\documentclass[preprint]{aastex63}

\usepackage{graphicx}
\usepackage{graphics}
\usepackage{hyperref}
\usepackage{bookmark}
\usepackage{float}
\usepackage{amsmath}
\DeclareMathOperator{\arctanh}{arctanh}

\shorttitle{Accelerated Expansion of the Universe with Non-Uniform Pressure}
\shortauthors{Kopteva~E., Bormotova~I., Churilova~M., Stuchlik~Z.}

\begin{document}

\title{Accelerated Expansion of the Universe in the Model with Non-Uniform Pressure}

\correspondingauthor{Elena Kopteva}
\email{olena.koptieva@gmail.com}

\author{Elena Kopteva}
\affiliation{Institute of Physics and Research Centre of Theoretical Physics and Astrophysics,\\ Faculty of Philosophy and Science in Opava, Silesian University in Opava, \\
746~01~Opava, Czech Republic} 
\author{Irina Bormotova}
\affiliation{Institute of Physics and Research Centre of Theoretical Physics and Astrophysics,\\ Faculty of Philosophy and Science in Opava, Silesian University in Opava, \\
746~01~Opava, Czech Republic} 
\affiliation{Bogoliubov Laboratory of Theoretical Physics, Joint Institute for Nuclear Research \\
141980~Dubna, Russia}
\author{Mariia Churilova}
\affiliation{Institute of Physics and Research Centre of Theoretical Physics and Astrophysics,\\ Faculty of Philosophy and Science in Opava, Silesian University in Opava, \\
746~01~Opava, Czech Republic} 
\author{Zdenek Stuchlik}
\affiliation{Institute of Physics and Research Centre of Theoretical Physics and Astrophysics,\\ Faculty of Philosophy and Science in Opava, Silesian University in Opava, \\
746~01~Opava, Czech Republic} 



\begin{abstract}
We present the particular case of the Stephani solution for shear-free perfect fluid with uniform energy density and non-uniform pressure. Such models appeared as possible alternative to the consideration of the exotic forms of matter like dark energy that would cause the acceleration of the universe expansion. These models are characterised by the spatial curvature depending on time. We analyze the properties of the cosmological model obtained on the basis of exact solution of the Stephani class and adopt it to the recent observational data. The spatial geometry of the model is investigated. We show that despite possible singularities, the model can describe the current stage of the universe evolution.
\end{abstract}



\section{Introduction}

Although $\Lambda$CDM model, based on the Friedmann solution, is most popular for explanation of the observed cosmological acceleration, it faces some fundamental problems, like the problem of ``dark energy'' and the coincidence problem \citep{Weinberg89}. Thus different attempts to find a possible alternative in this regard arise. The consideration of the inhomogeneous cosmological models is among them. The Stephani solution \citep{Stephani67} has drawn attention of cosmologists so long as it allows to build the model of the universe with accelerated expansion \citep{Dabrowski98,  Stelmach04, Stelmach08, Balcerzak15, Ong18}. This is a non-static solution for the expanding perfect fluid with zero shear and rotation, which contains the known Friedmann solution as a particular case. The Stephani solution was discussed much in the literature (see e.g. \citep{Krasinski83, Sussman87, Sussman88a, Sussman88b, Sussman00, Dabrowski93, Korkina16, Ong18} and references therein). Originally it has no symmetries, but the special case of spherical symmetry is of particular interest in cosmology. It is known that spatial sections of the Stephani space-time in this case have the same geometry as if they were subspaces of the Friedmann solution. Therefore these models may have an intuitively clear interpretation being in close connection with the Friedmann ones. The spatial curvature in the Stephani cosmological models is arbitrary function of time. This very property allows to obtain the appropriate behaviour of the cosmological acceleration.

According to our knowledge only a few of the cosmological models  based on the Stephani solution were studied concerning their correspondence to the observational data. 

In present work, we consider a rather general case of this solution restricted by the choice of the energy density in the same form as for the Friedmann dust. We analyze the properties of the resulting cosmological model and its applicability to the description of the current stage of the universe evolution. 
	
The paper is organized as follows. In Sec.~\ref{Sec2} we introduce the special case of the Stephani solution for our model and fit it to the current values of the cosmological parameters. The geometry of the spatial part of the obtained solution is explored in Sec.~\ref{Sec4}. In Sec.~\ref{Sec3} we investigate the dynamics of the universe evolution in our model, build the R-T-regions for the resulting space-time and discuss singularities of the model. In Sec.~\ref{Sec5} we consider the cosmological implications of our model. The conclusions are presented in Sec.~\ref{Con}.

\section{Special case of the Stephani solution}\label{Sec2}

It is known that for the perfect fluid described by 4-velocity vector field $u^\alpha$ there exist four main kinematic characteristics (see e.g. \citep{Stephani03}), which are the acceleration $\dot{u}_{\alpha}$,  the volume expansion $\Theta$, the rotation $\omega_{\alpha\beta}$ and the shear $\sigma_{\alpha\beta}$. These parameters are defined as follows:
\begin{eqnarray}\label{thetadef}
&\Theta &=u^{\alpha}_{;\alpha}, \\ 
&\omega_{\alpha\beta}&=u_{[\alpha;\beta]}+\dot{u}_{[\alpha}u_{\beta]}, \\
&\sigma_{\alpha\beta}&=u_{(\alpha;\beta)}+\dot{u}_{(\alpha}u_{\beta)}-\Theta h_{\alpha\beta}/3,\\
&h_{\alpha\beta}&=g_{\alpha\beta}+u_{\alpha}u_{\beta},
\end{eqnarray}
where $h_{\alpha\beta}$ is the projection tensor, greek indexes run from 0 to 3, square/round brackets standardly mean antisymmetryzation/symmetryzation by corresponded indexes. Here and further  in the paper dot means the partial derivative with respect to time and the geometric units are used where $c \equiv 1$, 
$ 8\pi G \equiv 1$. 

The Stephani solution is the solution of the Einstein equations for the universe filled with perfect fluid with zero shear and rotation. It is usually written in comoving coordinates in which 4-velocity of fluid particles has the following components: $u^0=1/\sqrt{g_{00}}$, $u^i=0, \, i=1,2,3$. 
 
In commonly used notations the Stephani solution in the case of spherical symmetry has the form \citep{Krasinski83}
\begin{equation}\label{genm}
\mathrm{d}s^2=D^2\mathrm{d}t^2-\frac{R^2(t)}{V^2(t,\chi)}\left(\mathrm{d}\chi^2+\chi^2\mathrm{d}\sigma^2\right),
\end{equation}
where $\mathrm{d}\sigma^2$ is the usual metric on the unit 2-sphere and
\begin{eqnarray}
&V&=1+\frac{1}{4}k(t)\chi^2 \label{v} \\
&D&=F(t)\left(\frac{V}{R}\right)^{.}\left(\frac{V}{R}\right)^{-1}.
\end{eqnarray}
The energy density is uniform and given by
\begin{equation}\label{e}
\varepsilon(t)=3C^2(t).
\end{equation}
The function $k(t)$ is defined by the expression
\begin{equation}\label{k} 
k(t)=\left[C^2(t)-\frac{1}{F^2(t)}\right]R^2(t),
\end{equation}
and corresponds to the curvature parameter, which in the Friedmann solution is constant normalizable to $0,\pm1$. Here $C(t)$, $F(t)$, $R(t)$ are arbitrary functions.

By use of the following relations
\begin{equation}\label{rRV}
r(t,\chi)=\frac{R(t)}{V(t,\chi)},
\end{equation}
\begin{equation}\label{psi}
\psi(t)=\frac{1}{F(t)},
\end{equation}
\begin{equation}\label{Ra}
a(t)=R(t),
\end{equation}
\begin{equation}\label{zkR}
\zeta(t)=\frac{k(t)}{R^2(t)}.
\end{equation}
the metric (\ref{genm}) may be rewritten in the form
\begin{equation}\label{ourm}
\mathrm{d}s^2=\frac{\dot{r}^2}{r^2\psi^2}\mathrm{d}t^2-r^2\left(\mathrm{d}\chi^2+\chi^2\mathrm{d}\sigma^2\right),
\end{equation}
which we shall use in further consideration as more convenient for our purposes. Here
\begin{equation}\label{r}
r(t,\chi)=\frac{a(t)}{1+\frac{1}{4}\zeta(t)a^2(t)\chi^2}\\,
\end{equation}
\begin{equation}\label{epsilon}
\zeta(t)=\varepsilon(t)-\psi^2(t),
\end{equation}
where $\zeta(t)$, $\psi(t)$ and $\varepsilon(t)$ are arbitrary functions, $a(t)$ is the function related to the scale factor of the Friedmann solution. 

According to the Einstein equations the pressure is defined by the expression
\begin{equation}\label{pres}
p(t,\chi)=-\varepsilon(t)-\frac{\dot{\varepsilon}(t)r(t,\chi)}{3\dot{r}(t,\chi)}.
\end{equation}
It is clear that the pressure is non-uniform in the Stephani solution.

The function $\zeta(t)$ is the spatial curvature. It is easy to verify that the scalar curvature $R$ of the spatial sections $t=\mathrm{const}$ of the metric (\ref{ourm}) is $R=6\zeta(t)$ and the Kretschmann invariant in this 3-dimensional case is $K=R_{\alpha\beta\mu\nu}R^{\alpha\beta\mu\nu}=12\zeta^2(t)$. As far as $t=\mathrm{const}$, one obtains the subspace of everywhere constant curvature. It is from here that the identity for the spatial geometries of the Stephani and the Friedmann solutions follows.

Further in this article we sometimes omit the variable of the function provided it will not cause the confusion.

As it can be deduced from the following consideration, the equation (\ref{epsilon}) is a generalization of the known Friedmann equation
\begin{equation}\label{Fr}
\frac{\dot{a}(t)^2}{a(t)^2}+\frac{k}{a(t)^2}=\frac{1}{3}\varepsilon(t),
\end{equation}    
where $k=0,\pm 1$, the factor $1/3$ in (\ref{epsilon}) is left out because of the arbitrariness of $\varepsilon(t)$.

The function $\psi(t)$ turns out to be connected with the Hubble parameter $H$. In the case of inhomogeneous cosmological model, the definition of the Hubble parameter should be generalized as it depends both on the time and spatial position. We shall use the generalization introduced by \citep{Ellis}:
\begin{equation}\label{H}
H=\frac{1}{l}\frac{\mathrm{d}l}{\mathrm{d}\tau}=\frac{1}{3}\Theta,
\end{equation}
where $l$ is some ``representative'' length that corresponds to the scale factor $a(t)$ in the Friedmann models, and $\tau$ is the proper time given in the standard way by
\begin{equation}
\mathrm{d}\tau=\sqrt{g_{00}}\mathrm{d}t.
\end{equation}
Due to the definition of the comoving coordinates, we obtain for the metric (\ref{ourm}) from the expression (\ref{thetadef})
\begin{equation}\label{Theta}
\Theta=\frac{1}{2\sqrt{g_{00}}}\frac{3}{g_{11}}\frac{\partial}{\partial t}g_{11}=\frac{3}{2}\frac{r\psi}{\dot{r}}\frac{2r}{r^2}\dot{r}=3\psi.
\end{equation}
From (\ref{H}) and (\ref{Theta})it follows that
\begin{equation}\label{hp}
H=\psi.
\end{equation}
If $\zeta=0$, then (\ref{ourm}) is a parabolic type of the Friedmann solution and it follows from the eq.~(\ref{epsilon})  that the function $\psi^2$ attains the sense of the critical energy density $\varepsilon_{\mathrm{cr}}=3H^2$ which is also in accordance with (\ref{hp}). 

For the parabolic Friedmann solution we have
\begin{equation}
r=a(t), \quad g_{00}=\frac{\dot{r}^2}{r^2\psi^2}=1 \quad \Rightarrow \quad g_{00}=\frac{\dot{a}^2}{a^2\psi^2} =1,
\end{equation}
then for the function $\psi(t)$ it follows that
\begin{equation}\label{hpa}
\psi(t)=\frac{\dot{a}}{a}.
\end{equation}
Thus for the appropriate transition to the Friedmann limit in the metric (\ref{ourm}) one should choose $\psi(t)$ in the form (\ref{hpa}).

\subsection{The model of the universe with the accelerated expansion}
We now define our model of the universe with the accelerated expansion based on the mentioned particular case of the Stephani solution.

We suppose the universe to be filled everywhere with the expanding shear-free perfect fluid with uniform energy density $\varepsilon=\varepsilon(t)$ and non-uniform pressure $p=p(\chi,t)$.

Let us start from the general Stephani metric in comoving coordinates written with conformally flat spatial part
\begin{equation}
\label{ourm1}
\mathrm{d}s^2=\frac{\dot{r}^2 a^2}{r^2 \dot{a}^2}\mathrm{d}t^2-r^2\left(\mathrm{d}x^2+\mathrm{d}\sigma^2\right),
\end{equation}
where 
\begin{equation}\label{rotx}
r(t,x)=\frac{2a(t)e^x}{1+\zeta(t)a^2(t)e^{2x}}.
\end{equation}
As discussed above, the main equation that governs the evolution of the model reads
\begin{equation}\label{maineq}
\frac{\dot{a}^2}{a^2}+\zeta(t)=\varepsilon(t). 
\end{equation}
From here and (\ref{Fr}) it evidently follows that the Stephani models with $\zeta=\pm 1/a^2,0$ are the Friedmann models.

The appropriate choice of the spatial coordinate transformation brings the spatial part of the metric (\ref{ourm1}) to one of the following forms
\begin{eqnarray}
&\mathrm{d}l^2&=\frac{a^2}{\left[ \cos^2\frac{\chi}{2}+\zeta a^2 \sin^2\frac{\chi}{2}\right]^2}\left(\mathrm{d}\chi^2+\sin^2\chi\mathrm{d}\sigma^2\right), \quad e^x=\tan\frac{\chi}{2},\\
&\mathrm{d}l^2&=\frac{a^2}{\left[ \cosh^2\frac{\chi}{2}+\zeta a^2 \sinh^2\frac{\chi}{2}\right]^2}\left(\mathrm{d}\chi^2+\sinh^2\chi\mathrm{d}\sigma^2\right), \quad e^x=\tanh\frac{\chi}{2},\\
&\mathrm{d}l^2&=\frac{a^2}{\left[ 1+\zeta a^2 \frac{\chi^2}{4}\right]^2}\left(\mathrm{d}\chi^2+\chi^2\mathrm{d}\sigma^2\right), \quad e^x=\frac{\chi}{2}.
\label{2.7}
\end{eqnarray}
For our description we shall choose the case (\ref{2.7}) implying the space-time metric
\begin{equation}\label{ourm2}
\mathrm{d}s^2=\frac{\dot{r}^2 a^2}{r^2 \dot{a}^2}\mathrm{d}t^2-r^2\left(\mathrm{d}\chi^2+\chi^2\mathrm{d}\sigma^2\right),
\end{equation}
\begin{equation}\label{rfin}
r=\frac{a}{1+\zeta a^2 \left(\frac{\chi}{2}\right)^2}.
\end{equation}
The energy density is chosen the same as for the Friedmann dust: 
\begin{equation}\label{enden}
\varepsilon =\frac{a_0}{a^3},
\end{equation}
$a_0=\mathrm{const}=a(t_0)$, where $t_0$ corresponds to the current moment of time (our time).

We take the spatial curvature in the form
\begin{equation}\label{2.9}
\zeta=-|\beta|\frac{a_0^k}{a^{k+2}},
\end{equation}
where $k=\mathrm{const}, \beta=\mathrm{const}<0$, that means that the spatial curvature is negative everywhere in the universe. Such expression for the spatial curvature is induced by the form of the Friedmann equation (\ref{Fr}) which may contain the sum of energy densities of several non-interacting sources. In the Friedmann models the energy density for all known components of matter (including those with negative pressure) is expressed in terms of scale factor raised to the correspondent power.  

The models known in the literature are mostly the particular cases of (\ref{2.9}) with fixed value of $k$ (\citep{Dabrowski95}, \citep{Dabrowski98} Model I: $k = -1$, Model II: $k = 1$,  \citep{Stelmach08, Stelmach04}: $k = -1$, \citep{Ong18}, \citep{Gregoris19}: $k = -1$).

Slightly different discussion of the models with unfixed $k$ is presented in \citep{Sussman00, Sedigheh14} in the frame of investigation of the Stephani universes with physically meaningful equations of state of matter.

We carry out our consideration without fixing $k$ but figuring out the range of its values that correspond to the right behavior of the universe acceleration.

The pressure in the model according to (\ref{pres}), (\ref{enden}) and (\ref{2.9}) reads
\begin{equation}
p=\frac{a_0}{a^3} \frac{\left(\frac{\chi}{2}\right)^2|\beta|k}{\left(\frac{a}{a_0}\right)^k-\left(\frac{\chi}{2}\right)^2|\beta|(k+1)}
\end{equation}
We now express some cosmological parameters in terms of $\chi$ and $a(t)$, which will somehow parametrize the time coordinate. In this part we restore the dimensions as far as we are going to put the numeric parameters of the model ($a_0$, $\beta$, $k$) in accordance with the observational data. 
\begin{enumerate}
\item The Hubble parameter 
From (\ref{maineq}) one has
\begin{equation}\label{at}
\frac{\dot{a}}{a}=\left[\frac{a_0}{a^3}+|\beta|\frac{a_0^k}{a^{k+2}}\right]^{\frac{1}{2}}, 
\end{equation}
and hence
\begin{equation}\label{star}
H=c\sqrt{\frac{a_0}{a^3}+|\beta|\frac{a_0^k}{a^{k+2}}}.
\end{equation}
\item Matter density parameter 
\begin{equation}\label{Om}
\Omega_{\mathrm{m}}=\frac{\varepsilon}{\varepsilon_{\mathrm{cr}}}=\frac{a_0 c^2}{3 a^3 H^2} 
\end{equation}
\item The radius of the universe at present time $r_0$
\begin{equation}
r_0=\int_0^{\chi_0}\sqrt{-g_{11}}\mathrm{d}\chi=\int_0^{\chi_0}r(t_0,\chi)\mathrm{d}\chi
\end{equation}
\begin{equation}
r_0=\int_0^{\chi_0}\frac{a(t_0)}{1+\zeta(t_0) a^2(t_0)\left( \frac{\chi}{2}\right)^2 }\mathrm{d}\chi.
\end{equation}
Taking $\zeta(t)$ from (\ref{2.9}) one obtains
\begin{equation}\label{r0}
r_0=\frac{4a_0}{|\beta|} \arctanh (\frac{\chi_0}{2}|\beta|).
\end{equation}
Using this relation it is possible to find the value of the coordinate $\chi_0$ corresponding to the current size of the universe: $r(\chi_0,t_0)=r_0$.

\item Deceleration parameter $q$. 
We take the general definition of the deceleration parameter according to \citep{Ellis}
\begin{equation}
q=-\frac{1}{l}\frac{\mathrm{d}^2 l}{\mathrm{d}\tau^2}\frac{1}{H^2}.
\end{equation}
From (\ref{H}), (\ref{hp}) and (\ref{hpa}) we have 
\begin{equation}\label{qe}
 \frac{\mathrm{d}l}{\mathrm{d}\tau}=l \frac{\dot{a}}{a}
\end{equation}
differentiating both parts of (\ref{qe}) with respect to the time we obtain 
\begin{equation}
\frac{\mathrm{d}}{\mathrm{d}\tau} \frac{\mathrm{d}l}{\mathrm{d}\tau}= \frac{\dot{a}}{a} \frac{\mathrm{d}l}{\mathrm{d}\tau} + \frac{l}{\sqrt{g_{00}}}  \frac{\mathrm{d}}{\mathrm{d}t} \left(\frac{\dot{a}}{a}\right) =H^2+\frac{r\dot{a}}{\dot{r}a}\left( \frac{\ddot{a}}{a} - H^2 \right)
\end{equation}
Finally for the deceleration parameter there is
\begin{equation}\label{decel}
q=-\left[1+ \frac{r\dot{a}}{\dot{r}a}\left( \frac{\ddot{a} a}{\dot{a}^2} - 1 \right) \right],
\end{equation}
or in the explicit form due to (\ref{star}):
\begin{equation}\label{qq}
q=\frac{k |\beta|\left( \frac{\chi}{2} \right)^2 }{\left( \frac{a}{a_0} \right)^k-(k+1)|\beta| \left( \frac{\chi}{2} \right)^2} \left[ 1 + \frac{ \left( \frac{a}{a_0} \right)^{k-1} +k|\beta| }{2\left(\left( \frac{a}{a_0} \right)^{k-1} + |\beta| \right)} \right].
\end{equation}
\end{enumerate}

It is clear that the expression (\ref{decel}) in the Friedmann limit ($r=a$) turns to the right form for the deceleration parameter in Friedman models: $q=-\frac{\ddot{a} a}{\dot{a}^2}$. 

\subsection{Estimation of the model constants with respect to the observational data}
In this subsection we introduce the comparison of some observable parameters from previous subsection with their values obtained within standard cosmological model. 
   
The current values of cosmological parameters obtained within $\Lambda$CDM model (or FLRW model with nonzero curvature) may be found in \citep{WMAP}. We shall assume the following numbers:
\begin{equation}\label{2.17}
H_0=2 \times 10^{-18}\mathrm{s}^{-1}, \quad \Omega_{\mathrm{m}}=0.3, \quad r_0 \approx 4,4 \times 10^{26} \mathrm{m}. 
\end{equation}
According to this data due to (\ref{star}) and (\ref{Om}), the constants of our model ($a_0, \, \beta, \, \chi_0$) related to the current moment of time can be defined as follows:\\
\begin{equation}\label{H0}
H_0=\frac{c}{a_0}\sqrt{1+|\beta|},
\end{equation}
\begin{equation}
\Omega_{\mathrm{m}}=\frac{c^2}{3a_0^2H_0^2},
\end{equation}
\begin{equation}
a_0=1.58\times 10^{26} \mathrm{m},
\end{equation}
\begin{equation}
\beta = -0.111113,
\end{equation}
\begin{equation}
\chi_0=2.59906.
\end{equation} 

\subsection{Singularities of the model}
It was also widely discussed \citep{Sussman88b, Krasinski83, Dabrowski98} that the Stephani models contain some special singularities that should be taken into account if one intends to build a cosmological model.
In our case the model contains three true singularities.  
\begin{enumerate}
\item The initial singularity: $a(t)=0$ $\Rightarrow$ $r=0$, $\varepsilon \to \infty$, $p \to \infty$.
\item The singularity arising from $g_{11}$:
\begin{equation}\label{sing}
\chi=\frac{2\left( \frac{a}{a_0} \right)^{\frac{k}{2}}}{\sqrt{|\beta|}}.
\end{equation}
\item The singularity arising from the expression for pressure $p$:
\begin{equation}\label{sing2}
\chi=\frac{2\left( \frac{a}{a_0} \right)^{\frac{k}{2}}}{\sqrt{(1+k)|\beta|}}.
\end{equation}
In the case of $k=-1$ the singularity points (\ref{sing2}) belong to the spatial infinity independently on the value of the time coordinate. This particular case is called in the literature the Stephani-Dabrowski model \citep{Dabrowski93, Stelmach08, Stelmach04}. If one chooses here $k<-1$ then the singular behavior of the pressure will disappear.
\end{enumerate}

\subsection{Mass function and horizons of the model}
Let us first briefly introduce the notion of R- and T-regions of the spherically symmetric space-time \citep{Novikov}.

The spherically symmetric metric written in general form
\begin{equation} \label{gm}
\mathrm{d}s^2=e^{\nu(t,x)}\mathrm{d}t^2-e^{\lambda(t,x)}\mathrm{d}x^2-r^2(t,x)\mathrm{d}\sigma^2
\end{equation}
can locally be brought to the view
\begin{equation} \label{cc}
\mathrm{d}s^2=A(\tilde{t},\tilde{x})\mathrm{d}\tilde{t}^2-B(\tilde{t},\tilde{x})\mathrm{d}\tilde{x}^2-\tilde{x}^2\mathrm{d}\sigma^2
\end{equation}
by coordinate transformation preserving the spherical symmetry:
\begin{equation} \label{trans}
\tilde{t}=\tilde{t}(t,x), \quad \tilde{x}=\tilde{x}(t,x).
\end{equation}
At the vicinity of a taken point two main situations are possible. First one is the case when the world line $\tilde{x}=\mathrm{const}$, $\theta=\mathrm{const}$, $\varphi=\mathrm{const}$ is time-like. In this case $\tilde{x}$ is the spatial coordinate, and the following inequality holds for the general metric (\ref{gm}) 
\begin{equation}\label{R-points}
e^{\nu-\lambda}>\left(\frac{\mathrm{d}x}{\mathrm{d}t}\right)^2.
\end{equation}
Here $\mathrm{d}x/\mathrm{d}t$ is found from the equations $\tilde{x}^2=r^2(t,x)=\mathrm{const}$, regarding the invariance of $g_{22}$ and $g_{33}$ under the transformation (\ref{trans}). 
The points for which the inequality (\ref{R-points}) is satisfied are called R-points. They form the R-region of the space-time with usual properties of the world and observers.

The second case is when the world line $\tilde{x}=\mathrm{const}$, $\theta=\mathrm{const}$, $\varphi=\mathrm{const}$ is space-like. In this case $\tilde{x}$ cannot be the spatial coordinate, thus in the metric (\ref{cc}), coordinates $\tilde{x}$ and $\tilde{t}$ ``change'' their roles (it is implied, that the functions $A(\tilde{t},\tilde{x})$ and $B(\tilde{t},\tilde{x})$ have the needed signs). In this case the following inequality holds for the general metric (\ref{gm})
\begin{equation}\label{T-points}
e^{\nu-\lambda}<\left(\frac{\mathrm{d}x}{\mathrm{d}t}\right)^2.
\end{equation} 
The points for which the inequality (\ref{T-points}) is satisfied are called T-points. They form the T-region of essential instability where static observer is impossible.
 
The strict equality 
\begin{equation}\label{hor}
e^{\nu-\lambda}=\left(\frac{\mathrm{d}x}{\mathrm{d}t}\right)^2
\end{equation}
defines the boundary between R- and T-regions of the space-time, known as horizon. 

Regarding the condition $\tilde{x}^2=r^2(t,x)=\mathrm{const}$ we rewrite (\ref{hor}) as follows
\begin{equation}\label{hor1}
e^{-\nu}\dot{r}^2=e^{-\lambda}r'^{2}.
\end{equation} 
This will be referred to as the horizon equation. The prime here means the partial derivative with respect to the spatial coordinate $x$.

The coordinate condition (\ref{hor1}) may also be expressed in terms of the so-called mass function \citep{KorkinaKopteva}, which for the metric (\ref{gm}) reads
\begin{equation}
m=r\left( 1+ e^{-\nu}\dot{r}^2-e^{-\lambda}r'^{2}\right).
\end{equation}

The horizon equation then transforms to
\begin{equation}\label{heq}
m=r.
\end{equation}
For the metric (\ref{ourm1}), regarding (\ref{rfin}) and (\ref{at}), the mass function takes the form
\begin{equation}\label{mff}
m=\frac{ a_0 \chi^3}{\left( 1-\left(\frac{a0}{a(t)}\right)^k \left(\frac{\chi}{2}\right)^2 |\beta| \right)^3}.
\end{equation}
The horizon equation (\ref{heq}) then gives the following expressions for two branches of the horizon
\begin{equation}\label{hor}
\chi_{1,2}=\frac{2}{|\beta|} \sqrt{\left( \frac{a(t)}{a_0}\right)^k \left[ 2\left( \frac{a(t)}{a_0}\right)^{k-1} + |\beta| \pm 2\left( \frac{a(t)}{a_0}\right)^{\frac{k-1}{2}}\sqrt{\left( \frac{a(t)}{a_0}\right)^{k-1} + |\beta|}     \right]}.
\end{equation}

\section{Geometry}\label{Sec4}
\begin{figure}[t]
\centering
	\begin{minipage}{0.6\linewidth}
 		\includegraphics[width=\linewidth]{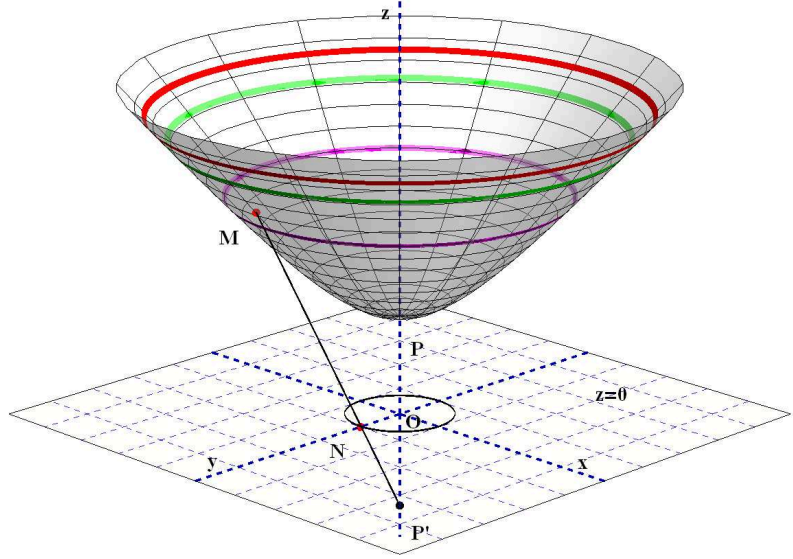}
	\end{minipage}	
\caption{The spatial section $t=t_0$, $\theta=\pi/2$ of the space-time (\ref{ourm2}). Origin \textbf{O} is the centre of the pseudo-sphere. Points \textbf{P} and \textbf{P$^\prime$} are the north and the south poles, respectively. The point \textbf{N} is the stereographic projection of the point \textbf{M}. The observer is situated at the point \textbf{P}.  The red line is the line $r=r_0$ indicating the current size of the universe. The green line is one branch of the horizon, corresponding to the sign ``$-$" in (\ref{hor}). The purple line is the so-called Hubble sphere, that expands with speed of light.}
\label{fig3}
\end{figure}

In this section we investigate the spatial geometry of the obtained solution. To build the spatial sections of the space-time with metric (\ref{ourm2}) we fix the time at present moment $t=t_0=\mathrm{const}$ that yields $a=a_0=\mathrm{const}$ in the formulae. To make it possible to visualize the 3-dimensional hypersurface we also fix $\theta=\pi/2$. Applying these conditions to (\ref{ourm2}) we obtain the intrinsic metric of the hypersurface of our interest in the following form
\begin{equation}\label{dl1}
\mathrm{d}l^2=\frac{a_0^2}{\left(1-\frac{|\beta|}{4}\chi^2\right)^2} \left(\mathrm{d}\chi^2+\chi^2 \mathrm{d}\varphi^2\right).
\end{equation}
By use of a new coordinate $\rho=\frac{\sqrt{|\beta|}}{2}\chi$ the metric (\ref{dl1}) can be rewritten in more familiar way
\begin{equation}\label{dl2}
\mathrm{d}l^2=\frac{a_0^2}{|\beta|}\frac{4}{\left(1-\rho^2\right)^2} \left(\mathrm{d}\rho^2+\rho^2 \mathrm{d}\varphi^2\right).
\end{equation}
This is a metric of the pseudo-sphere in terms of the stereographic projection coordinates (see e.g. \citep{Dubrovin}) accurate within the similarity transformation with constant factor $a_0^2/|\beta|$. This stereographic projection maps the upper half of the pseudo-sphere represented by the hyperboloid of revolution onto the open disk $\rho^2=x^2 + y^2 < 1$ on the plane $z=0$ as shown at Figure~\ref{fig3}. Such a disk equipped with normalized metric (\ref{dl2}) (so that $a_0^2/|\beta|=1$) refers to the Poincare model of Lobachevsky geometry. It is seen that the spatial sections of the interval (\ref{ourm2}) are the Lobachevsky spaces.

Figure~\ref{fig3} demonstrates the form of the spatial hypersurface of the universe within our model as an instantaneous snapshot at present moment of time, corresponding to the line $T=1$ at Fig.~\ref{fig1},\ref{fig2}. To restore the 3-dimensional picture from Fig.~\ref{fig3} one should imagine that the circles of the sections $z=\mathrm{const}$ are in fact 2-spheres.

\section{Universe evolution in the model}\label{Sec3}
To investigate the evolution of the universe in the obtained model we consider an observer situated close to the symmetry center $\chi=0$, who observes the dynamics of the infinite number of the concentric spheres marked by successive values of $\chi$.
The velocity of the expansion of some sphere $\chi$ may be found as follows
\begin{equation}\label{vel}
v(t,\chi)=\frac{\mathrm{d}}{\mathrm{d}t}\int_0^{\chi}r(t,\chi)\mathrm{d}\chi.
\end{equation}
Using the results obtained in Sec.~\ref{Sec2}, we now build the universe expansion velocity profile found from (\ref{vel}) and the deceleration parameter given by (\ref{qq}).

Further in our discussion we shall use the dimensionless function
\begin{equation}\label{T}
T \equiv \frac{a(t)}{a_0},
\end{equation} which will be treated as time parameter. The differentiation with respect to the time will be carried out taking into account that according to (\ref{at})
\begin{equation}\label{Tt}
\frac{\mathrm{d}T}{\mathrm{d}t}=\frac{1}{a_0}{\sqrt{\frac{1}{T}+\frac{|\beta|}{T^k}}}.
\end{equation}   
\begin{figure}[t]
	\begin{minipage}{0.49\linewidth}
 		\includegraphics[width=\linewidth]{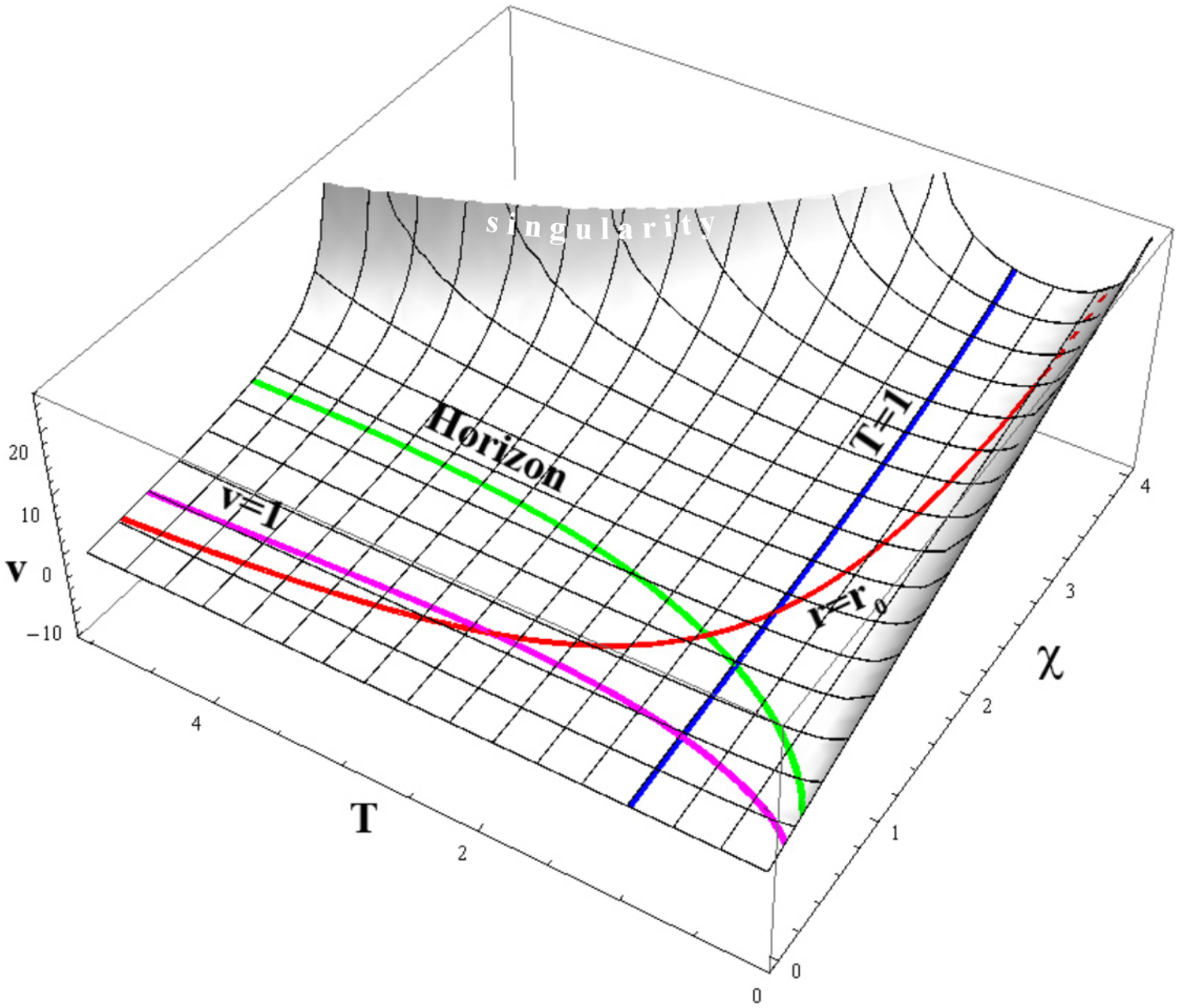}
	\end{minipage}
	\hspace{0.0005\linewidth}
 	\begin{minipage}{0.49\linewidth}
		\includegraphics[width=\linewidth]{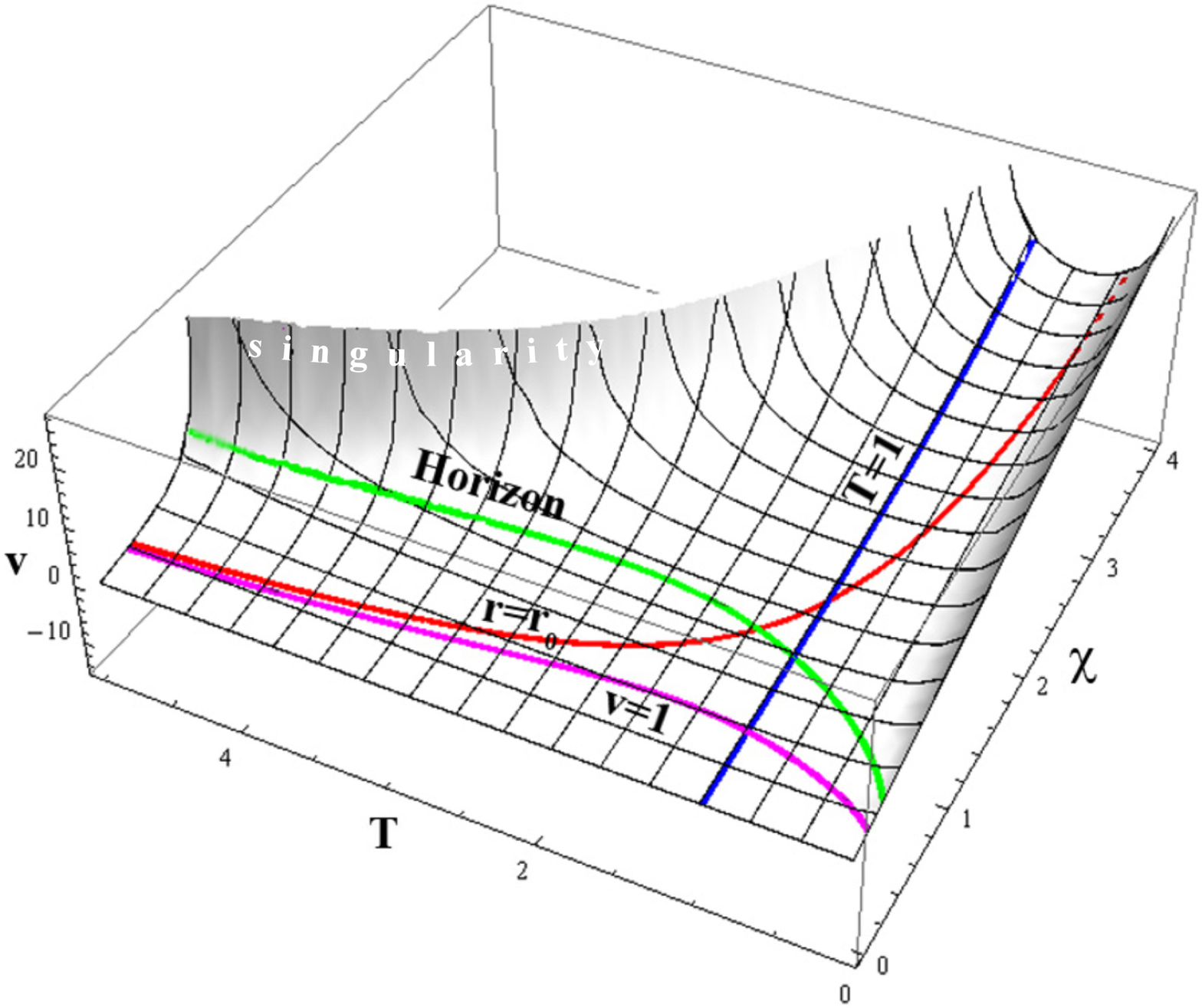}
	\end{minipage}
\caption{The universe expansion velocity profile for $k=-1.2$ (left side) and $k=-2.2$ (right side) in dimensionless units. The observer is situated in the center $\chi=0$. The line $T=1$ corresponds to the current moment of time $t=t_0$. The line $r=r_0$ is the line along which the radius of the universe equals to its current size found from the observations. The line $\mathrm{v}=1$ defines the Hubble sphere that expands with speed of light. The green line shows the ``visible'' branch of the horizon. The second branch lies behind the singularity. T-region of essential non-stationarity is situated between two branches of the horizon. R-region corresponding to our world, lies under the ``visible'' branch of the horizon.}
\label{fig1}
\end{figure}

Figure~\ref{fig1}  shows the universe expansion velocity profile in the model in terms of dimensionless units, where the time parameter $T$ is given by (\ref{T}). The concrete values of the index $k$ are chosen only for illustrative purposes, with decreasing of $k$ the picture qualitatively remains the same. The point of intersection of the lines $T=1$ and $r=r_0$ defines the coordinate $\chi_0$ that indicates the sphere of radius $r_0$ corresponding to the edge of the universe. It is seen that at present time the boundary of the universe belongs to the region of non-stationarity and expands with the velocity exceeding the speed of light as it is in standard Friedmann model. The central observer always belongs to the R-region of permitted observers. It is also seen that the universe does not reach the singularity given by (\ref{sing}) up to its present age, and the singularity cannot be observed according to the causality principle.

\begin{figure}[t]
	\begin{minipage}{0.49\linewidth}
 		\includegraphics[width=\linewidth]{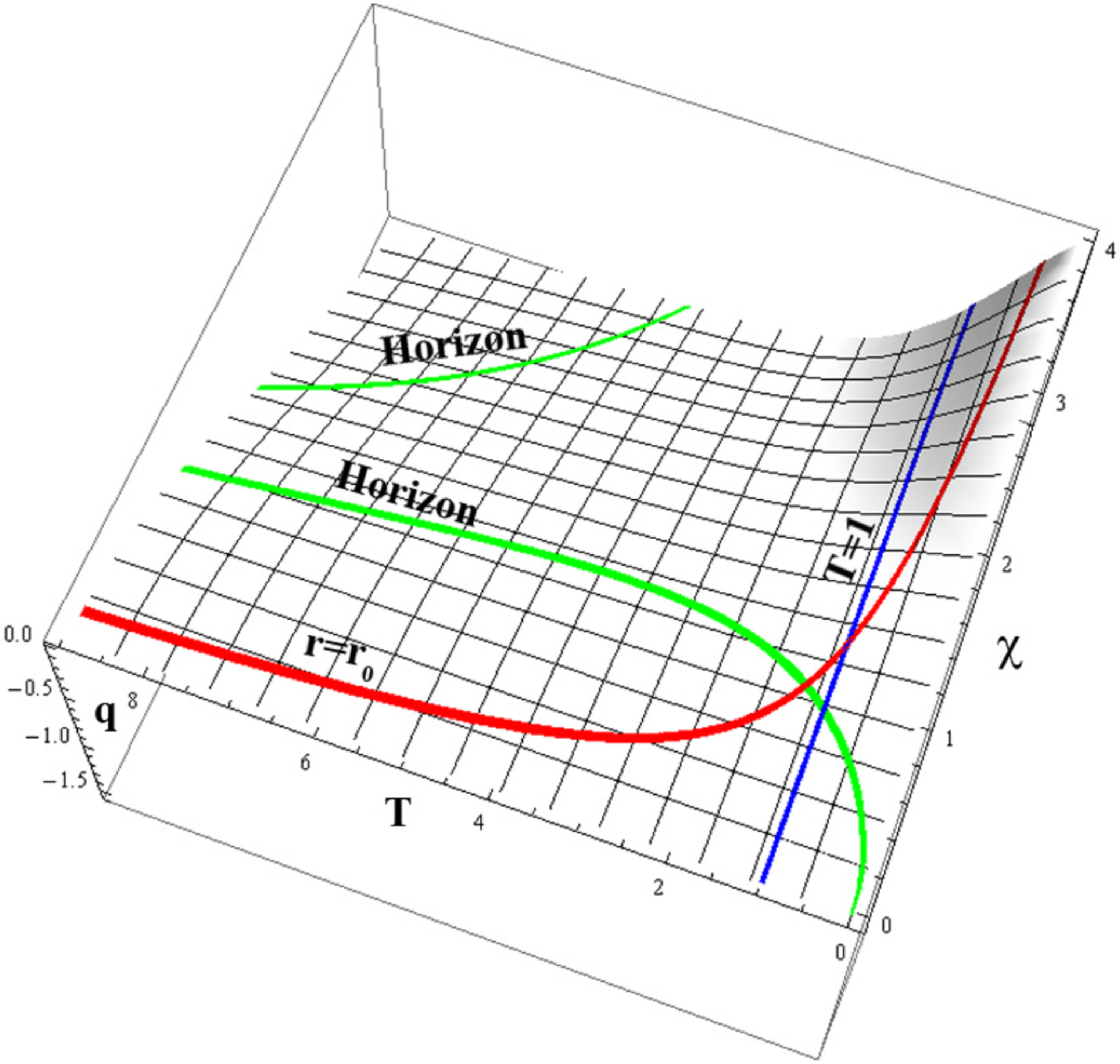}
	\end{minipage}
	\hspace{0.0005\linewidth}
 	\begin{minipage}{0.49\linewidth}
		\includegraphics[width=\linewidth]{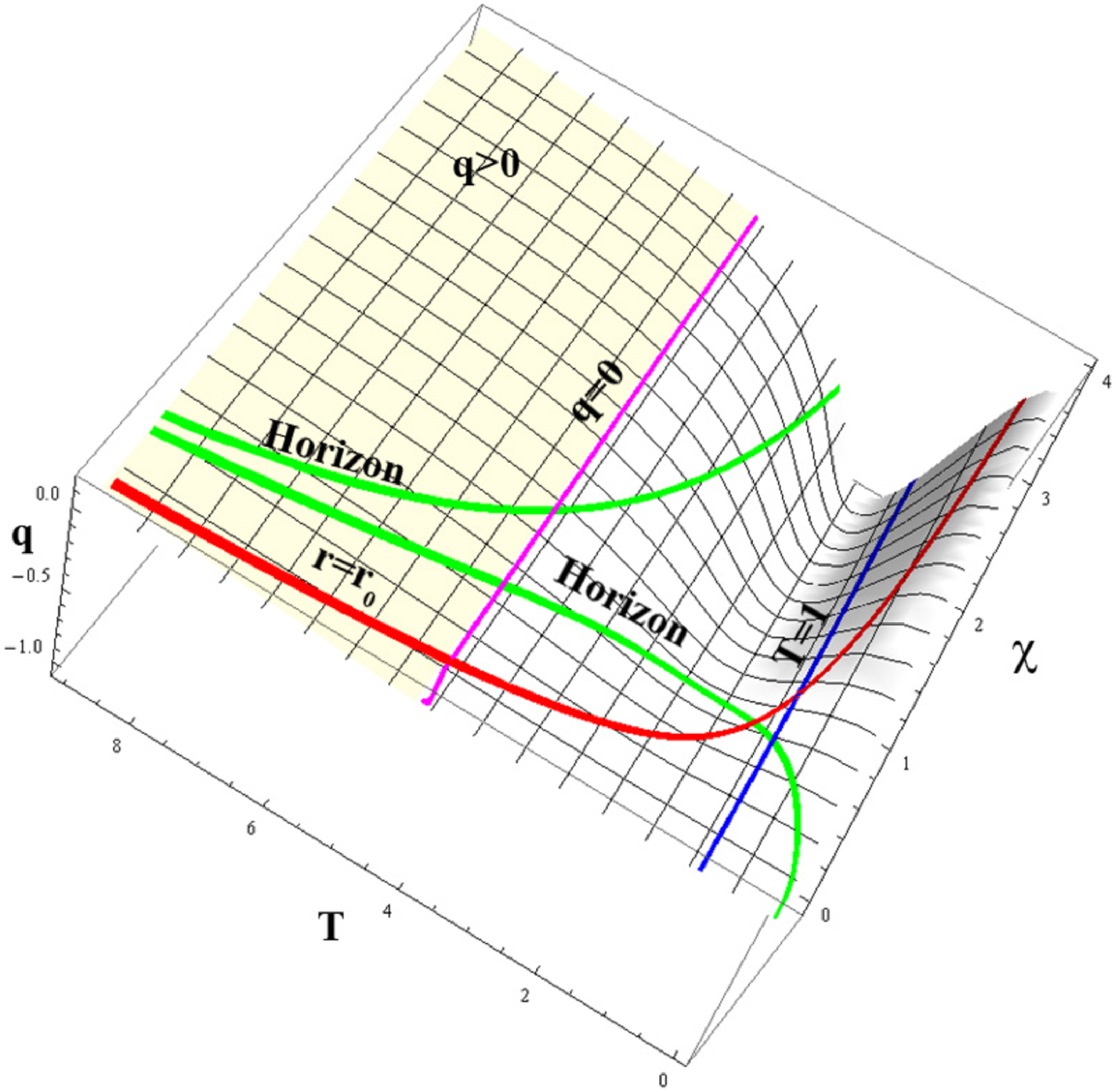}
	\end{minipage}
\caption{The profile of the deceleration parameter for $k=-1.2$ (left side) and $k=-2.2$ (right side). The line $T=1$ corresponds to the current moment of time. The green lines are two branches of the horizon. T-region is situated between the branches of the horizon. R-region is the region outlined by the lower branch of the horizon. The line $q=0$ is the line where the deceleration parameter equals to zero.}
\label{fig2}
\end{figure}

Figure~\ref{fig2} shows the behaviour of the deceleration parameter of the model. This profile is not affected by the singularity. For $k<-2$ there exists the line of zero deceleration parameter $q=0$. Hence one could expect that after some time the acceleration of the universe expansion changes into deceleration. However, even from the velocity profile (Fig.~\ref{fig1}) it is clear that there is no deceleration in the future. It may be verified by direct calculations using (\ref{vel}) that the function $\mathrm{d}v(t,\chi)/\mathrm{d}t$ changes its sign only once, from negative to positive. Thus, we conclude that in our model, unlike the Friedmann models, there is no correlation between the signs of the deceleration parameter and the acceleration of the universe expansion. 

\section{Redshift-magnitude relation in the model}\label{Sec5}
The advantage of the classical redshift-magnitude test is the sensitivity of the relation between the apparent magnitude and the redshift of the source to the cosmological model. In this section we compare the observational results concerning the redshift-magnitude relation for Supernovae type Ia with the theoretical predictions of our model. In this regard we shall use the Hubble diagram of distance moduli and redshifts for HST-discovered SNe Ia in the gold and silver sets represented at \citep{Riess04}.

The redshift-magnitude relation for the inhomogeneous Stephani model was derived first and studied in \citep{Dabrowski95},\citep{Dabrowski98} for two special cases near the observer position. Further, it was developed numerically by \citep{Stelmach08} for one of these cases, but higher redshifts.

For our model we derive the redshift-magnitude relation in terms of distance modulus analytically without any supposition about values of the redshift.  

The distance modulus within cosmological scales ($\mathrm{Mpc}$) is defined by (see e.g. \citep{Ellis})
\begin{equation}\label{mu-def}
\mu(z)=m(z)-M=5\log _{10}[d_L(z)/\mathrm{Mpc}]+25,
\end{equation}
where $z$ is the redshift, $m(z)$ is the apparent bolometric magnitude of a standard candle whose absolute bolometric magnitude is $M$; $d_L(z)$ is the luminosity distance to the source
\begin{equation}\label{ld-def}
d_L(z)=r_{c}(z)(1+z)^2.
\end{equation}
This relation holds rather general and does not depend on metric choice. Here $r_{c}(z)$ is comoving distance to the source by apparent size, which in our case is usual comoving radial distance given by \citep{Celerier2000}
\begin{equation}\label{rc}
r_{c}(t,\chi)=r(t,\chi)\chi=\frac{a_0 T \chi}{1-T^{-k}|\beta|\left(\frac{\chi }{2}\right)^2}
\end{equation} 
The general definition for the redshift in any cosmological model reads \citep{Ellis}
\begin{equation}\label{redshift-def}
1+z=\frac{(\kappa_\alpha u^\alpha)_{\mathrm{emitter}}}{(\kappa_\alpha u^\alpha)_{\mathrm{observer}}},
\end{equation}  
where $u^{\alpha}=\mathrm{d}x^{\alpha}/\mathrm{d}s$ is usual 4-velocity of the cosmological medium and  $\kappa^{\alpha}=\mathrm{d}x^{\alpha}/\mathrm{d}\lambda$ is a vector tangent to the correspondent null-geodesic with affine parameter $\lambda$, i.e. the solution of the geodesic equations for the photon. Indexes 'emitter' and 'observer' mean that the quantity should be calculated at the correspondent position.

Applying these definitions to the interval (\ref{ourm2}) we shall act according to the following plan:
\begin{enumerate}
\item Solving the geodesic equations for the photon radial motion we obtain $\kappa^\alpha$.
\item Taking into account the fact that in comoving system the only nonzero component of the 4-velocity is $u^0=1/\sqrt{g_{00}}$, we find the expression for the redshift in terms of the time and spatial coordinate.
\item Using previous results, we compose the distance modulus $\mu$ as a function of $T$ and $\chi$, according to (\ref{mu-def})-(\ref{rc}). Thus we obtain the two-parametric area $\mu-z$ with $T$ and $\chi$ being the parameters.
\item Then we use the condition that for any time $t$ there exists only one possible coordinate $\chi$ such that the light being emitted from the point $(t,\chi)$ will be received by the observer at the point $(t=t_0,\chi=0)$. This is expressed in the following equation
\begin{equation}\label{light}
\int_t^{t_0}\sqrt{g_{00}(t,\chi)}\mathrm{d}t=\int_0^{\chi}r(t,\chi)\mathrm{d}\chi,
\end{equation}
which in our notations (\ref{T}),(\ref{Tt}) gives
\begin{equation}\label{light2}
\int_T^{1}\frac{a_0\sqrt{g_{00}(T,\chi)}}{{\sqrt{T^{-1}+T^{-k}|\beta|}}}\mathrm{d}T=\int_0^{\chi}r(T,\chi)\mathrm{d}\chi.
\end{equation}
\item Then we plot the numerical solution of the equation (\ref{light2}) as the line $T(\chi)$ at the two-parametric diagram $\mu(T,\chi)-z(T,\chi)$. As a result we obtain the theoretical prediction for the redshift-magnitude relation in our model and compare it with observational data from \citep{Riess04}. 
\end{enumerate}

The geodesic equations for the interval (\ref{ourm2}) in case of the photon radial motion read
\begin{equation}\label{geqk1}
\frac{\mathrm{d}\kappa^1}{\mathrm{d}\lambda}=\frac{1}{2g_{11}}\left(-2\kappa^1\kappa^0\dot{g}_{11}-(\kappa^1)^2g'_{11}+(\kappa^0)^2g'_{00}\right),
\end{equation}
\begin{equation}\label{geqk0}
\frac{\mathrm{d}\kappa^0}{\mathrm{d}\lambda}=\frac{1}{2g_{00}}\left(-2\kappa^1\kappa^0g'_{00}+(\kappa^1)^2\dot{g}_{11}-(\kappa^0)^2\dot{g}_{00}\right),
\end{equation} 
\begin{equation}\label{geqint}
\kappa^0=\frac{\sqrt{-g_{11}}}{\sqrt{g_{00}}}\kappa^1.
\end{equation}
The metric coefficients in the interval (\ref{ourm2}) have the following explicit form:
\begin{equation}
g_{00}=\frac{\left(1-T^{-k}(1+k)|\beta|\left(\frac{\chi }{2}\right)^2 \right)^2}{\left(1-T^{-k}|\beta| \left(\frac{\chi }{2}\right)^2\right)^2},
\end{equation}
\begin{equation}
g_{11}=-\frac{\left(a_0 T\right)^2}{\left(1-T^{-k}|\beta|\left(\frac{\chi }{2}\right)^2\right)^2}.
\end{equation}
Putting $\kappa^0$ from (\ref{geqint}) into (\ref{geqk1}) and taking into account that  $\kappa^0=\mathrm{d}\chi/\mathrm{d}\lambda$, we obtain the differential equation with separable variables and integrate it with a result
\begin{equation}
\kappa^1=4 a_0^k T^k \frac{\left(1- T^{-k}|\beta|\left(\frac{\chi }{2}\right)^2 \right)^2}{1-T^{-k}(1+k)|\beta| \left(\frac{\chi }{2}\right)^2 }\left(\frac{1+\frac{\chi}{2} \sqrt{|\beta|} T^{-\frac{k}{2}}}{1-\frac{\chi}{2} \sqrt{|\beta| } T^{-\frac{k}{2}}}\right)^{\frac{2 T^{k/2}}{\sqrt{|\beta|}} \sqrt{T^{-1} + T^{-k}|\beta|}}, 
\end{equation}
and hence due to (\ref{geqint})
\begin{equation}
\kappa_0=g_{00}\kappa^0=-4 a_0^{1+k} T^{1+k} \left(\frac{1+\frac{\chi}{2} \sqrt{|\beta|} T^{-\frac{k}{2}}}{1-\frac{\chi}{2} \sqrt{|\beta| } T^{-\frac{k}{2}}}\right)^{\frac{2 T^{k/2}}{\sqrt{|\beta|}} \sqrt{T^{-1} + T^{-k}|\beta|}}. 
\end{equation}

\begin{figure}[t]
\centering
	\begin{minipage}{0.7\linewidth}
 		\includegraphics[width=\linewidth]{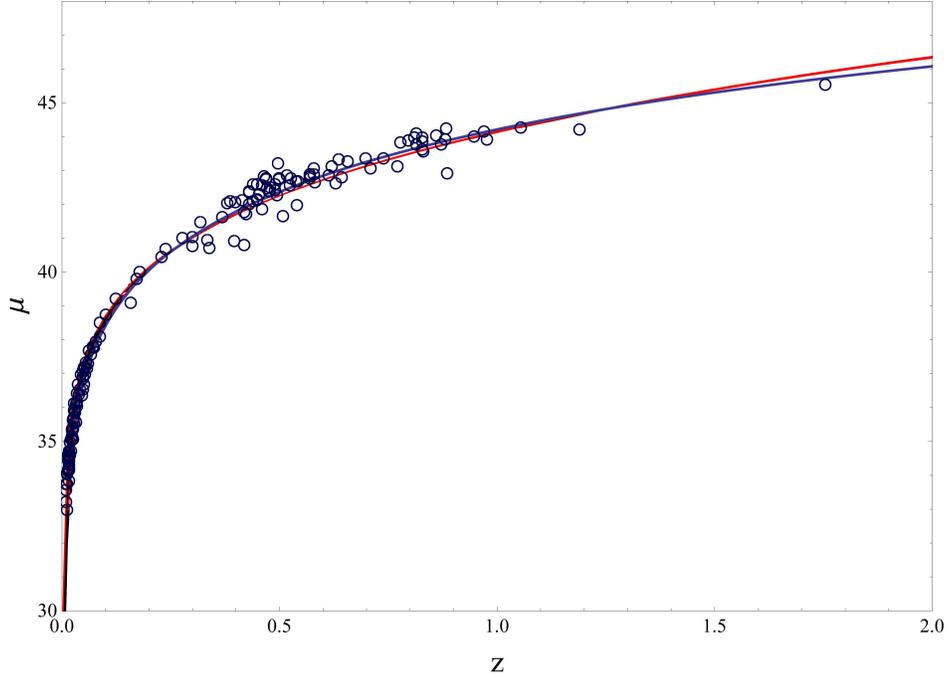}
	\end{minipage} 	
\caption{The redshift-magnitude relation in terms of distance modulus with overplotted data for HST-discovered SNe Ia in the gold and silver sets taken from \citep{Riess04}. The blue curve is the best-fit obtained by these authors in frame of the standard $\Lambda$CDM model with $H_0=66,9$ km/sec/Mpc, $\Omega_m=0.29$, $\Omega_{\Lambda}=0.71$. The red line is the best-fit found within the Stephani model with respect to the current time parameter $T=1.13$, curvature parameters: $\beta=-0.1$, $k=-1.01$, the Hubble constant current value $H_0=65.5642$ km/sec/Mpc.}
\label{fig4}
\end{figure}

In our model the observer occupies the position $\chi=0$ and receives the signal at present time $T=1$. Thus we obtain for the redshift
\begin{equation}\label{z}
1+z=\frac{\kappa_0u^0}{\kappa_0u^0\vert_{\chi=0,T=1}}= T^{1+k} \frac{1- T^{-k}|\beta|\left(\frac{\chi }{2}\right)^2}{1-T^{-k}(1+k)|\beta| \left(\frac{\chi }{2}\right)^2 }\left(\frac{1+\frac{\chi}{2} \sqrt{|\beta|} T^{-\frac{k}{2}}}{1-\frac{\chi}{2} \sqrt{|\beta| } T^{-\frac{k}{2}}}\right)^{\frac{2 T^{k/2}}{\sqrt{|\beta|}} \sqrt{T^{-1} + T^{-k}|\beta|}}.
\end{equation}
Now we have everything to plot the Hubble diagram for our model. The Fig.~\ref{fig4} shows the redshift-magnitude dependence in terms of the distance modulus (\ref{mu-def}) with overplotted data taken from \citep{Riess04}. It is seen that even in this particular case without concretizing the form of the function $a(t)$ the Stephani model can in principle give an adequate interpretation of the observational data. 

\section{Conclusions}\label{Con}
In this work, a particular case of the Stephani solution was investigated as possible model of the universe with accelerated expansion. The R-T-structure of the obtained space-time was built, and it was shown that the central observer belongs to the R-region of permitted observers. In this model the boundary of the observable universe belongs to the T-region and expands with velocity exceeding the speed of light, as it is in standard Friedmann model. The correlation between the signs of the deceleration parameter and the acceleration of the universe expansion is absent in this model. 

It is shown that the spatial sections of the universe are the Lobachevsky spaces. It turned out that the form of spatial section taken at present moment of time does not depend on the power of $a(t)$ in  curvature function $\zeta(t)$. 

It was established that the theoretical prediction for the redshift-magnitude relation in our model is in good accordance with type Ia Supernovae observational data.

The obtained results serve as an evidence in favour of the possibility that our world in principle may be described by such model up to its recent stage without any harm from existing singularities. Another advantage of this approach is that it allows to stay within the general relativity with no need for modifications and introducing any exotic types of matter.

\acknowledgments
This paper is supported by the Grant of the Plenipotentiary Representative of the Czech Republic in JINR under Contract No.~208 from 02/04/2019. The authors acknowledge the Research Centre of Theoretical Physics and Astrophysics of the Faculty of Philosophy and Science, Silesian University in Opava for support. Z.~S.\ acknowledges the Albert Einstein Centre for Gravitation and Astrophysics supported by the Czech Science Foundation Grant No.~14-37086G. I.~B.\ acknowledges the Silesian University in Opava grant SGS 12/2019. Authors cordially thank Maria Korkina for suggesting the problem and valuable discussion.


\begin{thebibliography}{}

\bibitem[Balcerzak, Dabrowski \& Denkiewicz (2015)]{Balcerzak15} Balcerzak, A., Dabrowski, M., \& Denkiewicz, T., \ 2015, \
\textit{A critical assessment of some inhomogeneous pressure Stephani models.} \ \href{https://arxiv.org/abs/1409.1523v3}{arXiv:1409.1523v3 [gr-qc]}

\bibitem[Celerier (2000)]{Celerier2000} Celerier, M.-N., \ 2000, \ Astron. \ Astrophys., { 353}, 63--71, 
\textit{Do we really see a cosmological constant in the supernovae data?} 
\href{https://arxiv.org/abs/astro-ph/9907206}{arXiv:astro-ph/9907206}

\bibitem[Dabrowski (1993)]{Dabrowski93} Dabrowski, M., \ 1993, \
J. \ Math. \ Phys. \ { 34}, 1447,
\textit{Isometric embedding of the spherically symmetric Stephani universe: Some explicit examples.} \href{https://doi.org/10.1063/1.530166}{DOI: 10.1063/1.530166}

\bibitem[Dabrowski (1995)]{Dabrowski95} Dabrowski, M., \ 1995, \
\ ApJ \ { 447}, 43,
\textit{A Redshift-Magnitude Relation for Non-Uniform Pressure
Universes.} \href{https://doi.org/10.1086/175855}{DOI: 10.1086/175855}

\bibitem[Dabrowski \& Hendry (1998)]{Dabrowski98} Dabrowski, M. \& Hendry M., \ 1998, \
\ ApJ \ { 498}, 67--76,
\textit{The Hubble Diagram of Type Ia Supernovae in Non-Uniform Pressure Universes.} \href{http://iopscience.iop.org/article/10.1086/305546}{DOI: 10.1086/305546}

\bibitem[Dubrovin (1992)]{Dubrovin} Dubrovin, B., Fomenko, A., Novikov, S. \ 1992, \ 
\textit{Modern Geometry - Methods and Applications. Part I. The Geometry of Surfaces, Transformation Groups, and Fields.} \ Springer-Verlag New York \ 

\bibitem[Ellis (2009)]{Ellis} Ellis, G.~F.~R., \ 2009, \
GRG \ { 41}, 581--660,
\textit{Republication of: Relativistic cosmology.} \href{https://doi.org/10.1007/s10714-009-0760-7}{DOI: 10.1007/s10714-009-0760-7}

\bibitem[Gregoris et al. (2019)]{Gregoris19} Gregoris, D., Ong, Y.~C., Wang B., \ 2019, \
\textit{Holographic Principle and the Second Law in Stephani Cosmology Revisited.} \href{https://arxiv.org/abs/1906.02879}{arXiv:1906.02879 [gr-qc]}

\bibitem[Hinshaw, G. et al. (2013)]{WMAP} Hinshaw, G. \ et \ al. \ 2013, \
Astrophys. \ J. \ Suppl. \ { 208}, 19,
\textit{Nine-Year Wilkinson Microwave Anisotropy Probe (WMAP) Observations: Cosmological Parameter Results.}
\href{https://arxiv.org/abs/1212.5226}{arXiv:1212.5226 [astro-ph.CO]}

\bibitem[Korkina \& Kopteva (2012)]{KorkinaKopteva} Korkina, M.~P. \& Kopteva, E.~M., \ 2012,
Space, \ Time \ and \ Fund. \ Interact. \ { 1}, 38--47, 
\textit{The mass function method for obtaining exact solutions in General Relativity.}
\href{https://arxiv.org/abs/1604.08247v2}{arXiv:1604.08247v2 [gr-qc]}

\bibitem[Korkina, Kopteva \& Egurnov (2016)]{Korkina16} Korkina, M.~P., Kopteva, E.~M. \& Egurnov, A.~A., \ 2016, \
Russ. \ Phys. \ J. \ { 59}, 3, 
\textit{Stephani Cosmological Models with Accelerated Expansion.} \href{https://doi.org/10.1007/s11182-016-0776-x}{DOI: 10.1007/s11182-016-0776-x}

\bibitem[Krasinski (1983)]{Krasinski83} Krasinski, A., \ 1983, \
GRG \ { 15}, 7
\textit{On the Global Geometry of the Stephani Universe.} \href{https://doi.org/10.1007/BF00759044}{DOI: 10.1007/BF00759044}

\bibitem[Misner, Thorne \& Wheeler (1973)]{Misner} Misner, C.~W., Thorne, K.~S. \& Wheeler J.~A., \ 1973,
\textit{Gravitation}, W.~H. Freeman and Company

\bibitem[Novikov (2001)]{Novikov} Novikov, I.~D., \ 2001,
\ Gen. \ Rel. \ Grav., { 33}, No. 12,
\textit{R- and T-Regions in Space-Time with Spherically Symmetric Space.} \href{https://doi.org/10.1023/A:1015398610011}{DOI: 10.1023/A:1015398610011}

\bibitem[Ong et al. (2018)]{Ong18} Ong, Y.~C., Hashemi, S.~S., An, R., Wang B., \ 2018, \
Eur. \ Phys. \ J. \ C { 78}, 405, 
\textit{Stephani cosmology: entropically viable but observationally challenged.} \href{https://link.springer.com/article/10.1140/epjc/s10052-018-5866-1}{DOI: 10.1140/epjc/s10052-018-5866-1}

\bibitem[Riess et al. (2004)]{Riess04} Riess, A. \ et \ al. \ 2004, \
Astrophys. \ J. \ { 607}, 2, 665--687,
\textit{Type Ia Supernova Discoveries at z > 1 from the Hubble Space Telescope: Evidence for Past Deceleration and Constraints on Dark Energy Evolution}
\href{https://iopscience.iop.org/article/10.1086/383612}{DOI: 10.1086/383612}

\bibitem[Hashemi et al. (2014)]{Sedigheh14} Hashemi, S.Sedigheh,  Jalalzadeh, S., Riazi, N., \ 2014, \ Eur. \ Phys. \ J. \ C { 74}, 2995,
 \textit{Dark side of the universe in the Stephani cosmology.} 
 \href{https://arxiv.org/abs/1401.2429}{arXiv:1401.2429 [gr-qc]}

\bibitem[Stelmach \& Szydlowski (2004)]{Stelmach04} Stelmach, J. \& Szydlowski, M., \ 2004, \
\textit{Can the Stephani model be an alternative to FRW accelerating models?}\ \href{https://arxiv.org/abs/astro-ph/0403534v1}{arXiv:astro-ph/0403534v1}

\bibitem[Stelmach \& Jakacka (2008)]{Stelmach08} Stelmach, J. \& Jakacka, I.,\ 2008, \
\textit{Nonhomogeneity driven Universe acceleration.} \ \href{https://arxiv.org/abs/0802.2284}{arXiv:0802.2284v1 [astro-ph]}

\bibitem[Stephani (1967)]{Stephani67} Stephani, H., \ 1967, \ 
Commun. \ Math.\ Phys. \ { 4}, 137--142  \href{https://projecteuclid.org/euclid.cmp/1103839813}{https://projecteuclid.org/euclid.cmp/1103839813}

\bibitem[Stephani et al. (2003)]{Stephani03} Stephani, H., Kramer, D. \ 2003, \ 
\textit{Exact solutions of Einstein's field equations.} \ Cambridge \ Univ. \ Press \

\bibitem[Sussman (1987)]{Sussman87} Sussman, R., \ 1987, \
J. \ Math. \ Phys. \ { 28}, 1118,
\textit{On spherically symmetric shear-free perfect fluid configurations (neutral and charged). I} \href{https://doi.org/10.1063/1.527558}{DOI: 10.1063/1.527558}

\bibitem[Sussman (1988a)]{Sussman88a} Sussman, R., \ 1988a, \
J. \ Math. \ Phys. \ { 29}, 945,
\textit{On spherically symmetric shear-free perfect fluid configurations (neutral and charged). II} \href{https://doi.org/10.1063/1.527992}{DOI: 10.1063/1.527992}

\bibitem[Sussman (1988b)]{Sussman88b} Sussman, R., \ 1988b, \
J. \ Math. \ Phys. \ { 29}, 1177,
\textit{On spherically symmetric shear-free perfect fluid configurations (neutral and charged). III} \href{https://doi.org/10.1063/1.527962}{DOI: 10.1063/1.527962}

\bibitem[Sussman (2000)]{Sussman00} Sussman, R., \ 2000, \
Gen. \ Rel. \ Grav. \ { 32}, 1527--1557, \textit{Towards a physical interpretation for the Stephani Universes}. \ \href{https://arxiv.org/abs/gr-qc/9908019}{arXiv:9908019 [gr-qc]}

\bibitem[Weinberg (1989)]{Weinberg89} Weinberg, S., \ 1989, \ 
Rev. \ Mod. \ Phys. \ { 61}, 1, \textit{The cosmological constant problem.}
           
           

\end{thebibliography}
\end{document}